\documentclass[aps,prl,reprint,groupedaddress,nofootinbib]{revtex4-1}
\usepackage[latin9]{inputenc}
\usepackage{latexsym,amssymb,amstext,amsmath}
\usepackage{slashed}
\usepackage{mathrsfs}
\usepackage{color}
\usepackage{hyperref}
\usepackage{verbatim}
\usepackage{graphicx}
\usepackage{color}
\usepackage{mathrsfs}
\begin{document}

\title{Schr\"{o}dinger connection with selfdual nonmetricity vector in 2+1 dimensions}
\author{Silke Klemm$^{1,2}$}
\author{Lucrezia Ravera$^{3,4}$}
\email{silke.klemm@mi.infn.it}
\email{lucrezia.ravera@polito.it}


\affiliation{$^1$ Dipartimento di Fisica, Universit\`{a} di Milano, Via Celoria 16, 20133 Milano, Italy}
\affiliation{$^2$ INFN, Sezione di Milano, Via Celoria 16, 20133 Milano, Italy}
\affiliation{$^3$ DISAT, Politecnico di Torino, Corso Duca degli Abruzzi 24, 10129 Torino, Italy}
\affiliation{$^4$ INFN, Sezione di Torino, Via P. Giuria 1, 10125 Torino, Italy}

\begin{abstract}
We present a three-dimensional metric affine theory of gravity whose field equations lead to a connection
introduced by Schr\"{o}dinger many decades ago. Although involving nonmetricity, the Schr\"{o}dinger 
connection preserves the length of vectors under parallel transport, and appears thus to be more
physical than the one proposed by Weyl. By considering solutions with constant scalar curvature, we obtain
a self-duality relation for the nonmetricity vector which implies a Proca equation that may also be
interpreted in terms of inhomogeneous Maxwell equations emerging from affine geometry.
\end{abstract}

\maketitle

\bigskip

\section{Introduction}

In 1918, Weyl proposed a remarkable generalization of Riemannian geometry (see
e.g.~\cite{Adler:1965,CP1,Folland,Romero:2012hs}) with an additional symmetry in an attempt of 
geometrically unifying electromagnetism with gravity \cite{Weyl:1918ib, Weyl:1919fi}. In this theory, both
the direction and the length of vectors vary under parallel transport. The connection introduced by Weyl 
involves a nonmetricity tensor whose trace part is known as the Weyl vector. However, Weyl's
attempt to identify the trace part of the nonmetricity, associated with stretching and contraction, with the
electromagnetic vector potential failed, due to observational inconsistencies \cite{Wheeler:2018rjb}.

On the other hand, in a series of papers written in the 1940s and collected in \cite{schrodinger}, with the
aim to construct a unified field theory, Schr\"{o}dinger proposed a symmetric connection which, although 
involving nonmetricity, preserves the length of vectors under parallel transport.
The Schr\"{o}dinger connection has the form
\begin{equation}\label{schconndef}
\hat{\Gamma}^\lambda_{\phantom{\lambda}\mu\nu} = \tilde{\Gamma}^\lambda_{\phantom{\lambda}
\mu\nu} + g^{\lambda\rho} Z_{\rho\mu\nu}\,,
\end{equation}
where $\tilde{\Gamma}^\lambda_{\phantom{\lambda}\mu\nu}$ denotes the Levi-Civita connection and 
$Z_{\mu\nu\rho}$ is a tensor obeying
\begin{equation}
Z_{\lambda\mu\nu} = Z_{\lambda\nu\mu}\,, \quad Z_{(\lambda\mu\nu)} = 0\,.
\end{equation}
The generic decomposition of an affine connection ${\Gamma^\lambda}_{\mu\nu}$ is given by
\begin{equation}\label{gendecconn}
{\Gamma^\lambda}_{\mu\nu} = \tilde{\Gamma}^\lambda_{\phantom{\lambda}\mu\nu} + 
\underbrace{{N^\lambda}_{\mu\nu}}_{\text{distortion}}\,,
\end{equation}
with
\begin{equation}\label{distortion3d}
\begin{split}
{N^\lambda}_{\mu\nu} & = \underbrace{\frac12 g^{\rho\lambda}\left(Q_{\mu\nu\rho} + Q_{\nu\rho\mu}
- Q_{\rho\mu\nu}\right)}_{\text{deflection}} \\
& -\underbrace{g^{\rho\lambda}\left(T_{\rho\mu\nu} + T_{\rho\nu\mu} - T_{\mu\nu\rho} 
\right)}_{\text{contorsion}}\,,
\end{split}
\end{equation}
where $Q_{\lambda\mu\nu}\equiv -\nabla_\lambda g_{\mu\nu} = -\partial_\lambda g_{\mu\nu} +
{\Gamma^\rho}_{\mu\lambda} g_{\rho\nu} + {\Gamma^\rho}_{\nu\lambda} g_{\mu\rho}$ and
${T_{\mu\nu}}^\lambda = {\Gamma^\lambda}_{[\mu\nu]}$ are the nonmetricity and the torsion tensor 
respectively \cite{Note1}.
In the case of vanishing torsion and $N_{(\lambda\mu\nu)}=0$, the connection \eqref{gendecconn}
reduces to
\begin{equation}\label{schnonmetr}
{\Gamma^\lambda}_{\mu\nu} = \tilde{\Gamma}^\lambda_{\phantom{\lambda}\mu\nu} - g^{\lambda\rho} 
Q_{\rho\mu\nu}\,,
\end{equation}
with
\begin{equation}\label{nmcomplsymmzero}
Q_{(\lambda\mu\nu)} = 0\,.
\end{equation}
One immediately sees that \eqref{schnonmetr} coincides with the Schr\"{o}dinger connection 
\eqref{schconndef} if we identify
\begin{equation}\label{zqident}
Z_{\lambda\mu\nu} = - Q_{\lambda\mu\nu}\,.
\end{equation}
Any connection respecting \eqref{nmcomplsymmzero} preserves the length of (but in general not the angle 
between) vectors under parallel transport \cite{schrodinger,Iosifidis:2019jgi}. Notice that, on the other hand, 
one could consider vanishing nonmetricity and then symmetrize the connection \eqref{gendecconn} in
$\mu,\nu$ to obtain
\begin{equation}\label{otherSch}
{\Gamma^\lambda}_{(\mu\nu)} := \check{\Gamma}^\lambda_{\phantom{\lambda}(\mu\nu)} = \tilde{\Gamma}^\lambda_{\phantom{\lambda}\mu\nu} - 2 g^{\lambda\rho} T_{\rho(\mu\nu)}\,.
\end{equation}
Then, $\check{\Gamma}^\lambda_{\phantom{\lambda}(\mu\nu)}$ coincides with \eqref{schconndef}
under the identification 
\begin{equation}\label{otheridT}
Z_{\lambda\mu\nu} = - 2 T_{\lambda (\mu\nu)}\,.
\end{equation}
Comparing \eqref{zqident} with \eqref{otheridT}, we see that \eqref{schconndef} can be written either in 
terms of torsion, in terms of nonmetricity fulfilling \eqref{nmcomplsymmzero}, or in terms of both of them.

The Schr\"{o}dinger connection seems to have been overlooked in the successive literature, despite its relevant features and the fact that it appears to be more physical than the one proposed by Weyl.
In this paper, we present a metric affine theory of gravity in 2+1 spacetime dimensions whose field
equations lead to a Schr\"{o}dinger connection.
Intriguingly, by considering solutions with constant scalar curvature, we obtain a self-duality
relation \cite{Townsend:1983xs} for the nonmetricity vector. This implies a Proca equation which may also
be interpreted in terms of inhomogeneous Maxwell equations emerging from affine geometry, i.e., from a 
purely gravitational setup. In this scenario, gauge invariance follows from self-duality and we can properly 
dub the nonmetricity vector `photon'.

\section{Schr\"{o}dinger connection with selfdual nonmetricity vector}\label{thy}

Before introducing our theory, let us recall the irreducible decomposition of the nonmetricity tensor
$Q_{\lambda\mu\nu}$ under the Lorentz group, that reads in three dimensions
\begin{equation}
\begin{split}
Q_{\lambda\mu\nu} & = \frac25 Q_{\lambda} g_{\mu\nu} - \frac15\tilde{Q}_\lambda g_{\mu\nu} +
\frac35 g_{\lambda(\nu}\tilde{Q}_{\mu)} \\
& -\frac15 g_{\lambda(\nu}Q_{\mu)} + \Omega_{\lambda\mu\nu}\,, \label{gennm}
\end{split}
\end{equation}
where $Q_\lambda\equiv {Q_{\lambda\mu}}^\mu$ and $\tilde{Q}_\lambda\equiv {Q^\mu}_{\mu\lambda}$ 
are nonmetricity vectors and $\Omega_{\lambda\mu\nu}$ is the traceless part of the nonmetricity.

We propose the action
\begin{equation}\label{actiontot}
\begin{split}
S & = \frac1{2\kappa^2}\int d^3 x\left(\sqrt{-g} f(R) + \frac1{2\mu} \epsilon^{\mu\nu\rho} Q_\rho 
\hat{R}_{\nu\mu}\right) \\
& + \int d^3 x\,\epsilon^{\mu\nu\rho}\zeta_{\nu\sigma} {T_{\rho\mu}}^\sigma\,,
\end{split}
\end{equation}
where $\kappa$ denotes the gravitational coupling constant, $f(R)$ is an arbitrary function of the scalar 
curvature $R = g^{\mu\nu} R_{\mu\nu}(\Gamma)$ ($\Gamma$ is a general affine connection), 
$\hat{R}_{\mu\nu} := {R^\lambda}_{\lambda\mu\nu} =
\partial_{[\mu}Q_{\nu]}$ denotes the homothetic curvature tensor, $\mu$ is a Chern-Simons
coupling constant, and $\zeta_{\nu\sigma}$ a Lagrange multiplier. In \eqref{actiontot} we also introduced
the Levi-Civita symbol $\epsilon^{\mu\nu\rho}=\sqrt{-g}\varepsilon^{\mu\nu\rho}$, where 
$\varepsilon^{\mu\nu\rho}$ is the Levi-Civita tensor. The action \eqref{actiontot} is manifestly
diffeomorphism-invariant.

We work in the Palatini formalism, where the metric $g_{\mu\nu}$ and the connection 
${\Gamma^\lambda}_{\mu\nu}$ are independent variables.
From the variation of \eqref{actiontot} with respect to $\zeta_{\mu\nu}$, we get vanishing torsion,
\begin{equation}\label{zerotor}
{T_{\rho\sigma}}^\nu = 0\,.
\end{equation}
The variation w.r.t~$g^{\mu\nu}$ leads to
\begin{equation}
f'(R) R_{(\mu\nu)} - \frac12 f(R) g_{\mu\nu} = 0\,. \label{eq:Einsteinnew}
\end{equation}
Notice that the Chern-Simons term and the piece involving the Lagrange multiplier do not contribute to \eqref{eq:Einsteinnew}. The trace of \eqref{eq:Einsteinnew} yields
\begin{equation}\label{trace3d}
\frac f{2f'} = \frac R3\,,
\end{equation}
which is identically satisfied if we choose
\begin{equation}\label{confchoice}
f(R) = C R^{3/2}\,,
\end{equation}
with $C$ an arbitrary integration constant. With the choice \eqref{confchoice}, the action \eqref{actiontot}
becomes invariant under the conformal transformation (see also \cite{Iosifidis:2019jgi,Klemm:2020mfp})
\begin{equation}\label{weylresc}
g_{\mu\nu}\mapsto g'_{\mu\nu} = e^{2\Omega} g_{\mu\nu}\,, \qquad
{\Gamma^\lambda}_{\mu\nu}\mapsto {\Gamma'^\lambda}_{\mu\nu} = {\Gamma^\lambda}_{\mu\nu}\,,
\end{equation}
where $\Omega$ is a scalar function.
On the other hand, \eqref{trace3d} can also be viewed as an algebraic equation for $R$ admitting
generically solutions with constant scalar curvature \cite{Iosifidis:2019jgi},
\begin{equation}\label{rconst}
R= c_k\,.
\end{equation}
Plugging \eqref{trace3d} into \eqref{eq:Einsteinnew}, the latter becomes
\begin{equation}
R_{(\mu\nu)} - \frac R3 g_{\mu\nu} = 0\,. \label{eq:EW3d}
\end{equation}
Observe that, together with \eqref{zerotor}, \eqref{eq:EW3d} would correspond to the Einstein-Weyl 
equations in three dimensions in the case in which one considers Weyl nonmetricity (see
e.g.~\cite{Klemm:2020mfp}). 
Varying \eqref{actiontot} w.r.t.~${\Gamma^\lambda}_{\mu\nu}$ and using \eqref{zerotor}, one gets
\begin{equation}\label{eqconnstart}
\begin{split}
& {P_\lambda}^{\mu\nu} + {\delta_\lambda}^\nu\frac{\partial^\mu f'}{f'} - g^{\mu\nu} 
\frac{\partial_\lambda f'}{f'} \\
& + \frac2{\mu f'}\varepsilon^{\nu\rho\sigma} {\delta_{\lambda}}^\mu \hat{R}_{\rho\sigma} + 
\frac{2\kappa^2}{f'}\varepsilon^{\mu\nu\rho}\zeta_{\rho\lambda} = 0\,,
\end{split}
\end{equation}
where
\begin{equation}
{P_{\lambda}}^{\mu\nu} = -\frac{\nabla_\lambda\left(\sqrt{-g} g^{\mu\nu}\right)}{\sqrt{-g}} + 
\frac{\nabla_\sigma\left(\sqrt{-g} g^{\mu\sigma}\right){\delta^\nu}_\lambda}{\sqrt{-g}}
\end{equation}
is the Palatini tensor with vanishing torsion.
Taking the $\lambda,\mu$ trace of \eqref{eqconnstart} and contracting with the Levi-Civita tensor, we get
\begin{equation}\label{lmtr}
\zeta_{[\rho\sigma]} = \frac3{\kappa^2\mu} \hat{R}_{\rho\sigma}\,.
\end{equation}
Plugging \eqref{lmtr} into \eqref{eqconnstart}, using \eqref{gennm}, and taking the $\lambda,\nu$ trace
of the resulting equation, one finds
\begin{equation}\label{nltr}
\tilde{Q}^\mu - \frac{Q^\mu}2 + \frac{\partial^\mu f'}{f'} - \frac4{\mu}\varepsilon^{\mu\rho\sigma} 
\hat{R}_{\rho\sigma} = 0\,.
\end{equation}
Then, with \eqref{lmtr} and \eqref{nltr}, the $\mu,\nu$ trace of \eqref{eqconnstart} yields
\begin{equation}\label{mntr}
\frac{\partial_\lambda f'}{f'} = \frac16 Q_\lambda + \frac2{3\mu f'}\varepsilon^{\alpha\rho\sigma}
g_{\alpha\lambda}\hat{R}_{\rho\sigma}\,.
\end{equation}
Inserting also \eqref{mntr} into \eqref{eqconnstart} after use of \eqref{lmtr} and \eqref{nltr}, and taking 
different contractions with the Levi-Civita tensor, we obtain
\begin{equation}\label{zetasymmzero}
\zeta_{(\lambda\nu)} = 0\,,
\end{equation}
and vanishing traceless part of the nonmetricity,
\begin{equation}\label{zeroOm}
\Omega_{\lambda\mu\nu} = 0\,.
\end{equation}
Using \eqref{lmtr} and \eqref{zetasymmzero}, one gets
\begin{equation}\label{zetaform}
\zeta_{\rho\sigma} = \frac3{\kappa^2\mu}\hat{R}_{\rho\sigma}\,.
\end{equation}
With \eqref{nltr}, \eqref{mntr}, \eqref{zeroOm} and \eqref{zetaform}, \eqref{eqconnstart} becomes
\begin{equation}\label{eqconn6}
\hat{R}_{\rho\sigma} {\delta_\lambda}^{[\mu}\varepsilon^{\nu]\rho\sigma} + \hat{R}_{\lambda\rho} 
\varepsilon^{\mu\nu\rho} = 0\,,
\end{equation}
Plugging \eqref{mntr} into \eqref{nltr}, we find
\begin{equation}\label{homotfin}
\hat{R}_{\rho\sigma} = \frac{\mu f'}{20}\varepsilon^{\mu\nu\lambda} g_{\rho\mu} g_{\sigma\nu}
\left(Q_\lambda - 3\tilde{Q}_\lambda\right)\,,
\end{equation}
which, used in \eqref{mntr}, leads to
\begin{equation}\label{dff}
\frac{\partial_\lambda f'}{f'} = \frac1{10}\left(Q_\lambda + 2\tilde{Q}_\lambda\right)\,.
\end{equation}
Notice also that, exploiting \eqref{homotfin}, \eqref{zetaform} becomes
\begin{equation}\label{zetaformfin}
\zeta_{\rho\sigma} = \frac{3 f'}{20 \kappa^2}\varepsilon^{\mu\nu\lambda} g_{\rho\mu} g_{\sigma\nu}
\left( Q_\lambda - 3\tilde{Q}_\lambda\right)\,.
\end{equation}
Finally, using \eqref{homotfin} we can see that \eqref{eqconn6} is identically satisfied.

Summarizing, one has
\begin{equation}\label{nmform}
Q_{\lambda\mu\nu} = \frac25 Q_{\lambda} g_{\mu\nu} - \frac15\tilde{Q}_\lambda g_{\mu\nu} + \frac35 
g_{\lambda(\nu}\tilde{Q}_{\mu)} - \frac15 g_{\lambda(\nu} Q_{\mu)}\,,
\end{equation} 
together with \eqref{zerotor}, \eqref{eq:EW3d}, \eqref{homotfin} and \eqref{dff}. The final form of the 
connection, obtained by plugging \eqref{zerotor} and \eqref{nmform} into \eqref{distortion3d}, results to be
\begin{equation}\label{connint}
\begin{split}
{\Gamma^\lambda}_{\mu\nu} & = \tilde{\Gamma}^\lambda_{\phantom{\lambda}\mu\nu} - \frac3{10} 
g_{\mu\nu} Q^\lambda + \frac25 {\delta_{(\mu}}^\lambda Q_{\nu)} \\
& + \frac25 g_{\mu\nu}\tilde{Q}^\lambda - \frac15 {\delta_{(\mu}}^\lambda\tilde{Q}_{\nu)}\,.
\end{split}
\end{equation}

Observe that with the choice \eqref{confchoice} (with $C=1$), \eqref{homotfin} and \eqref{dff} would
lead to the generalized monopole equation
\begin{equation}\label{genmoneq}
\star\left(d\Sigma + h\Sigma\right) = dh\,,
\end{equation}
where the one-form $h$ and the function $\Sigma$ are respectively defined by
$h_\lambda=-Q_\lambda/6$ and $\Sigma=3\mu\sqrt R/8$, together with
\begin{equation}\label{QQtrel}
\partial_\mu\ln R = \frac15\left(Q_\mu + 2\tilde{Q}_\mu\right)\,.
\end{equation}
Actually, \eqref{genmoneq} represents a special case of the generalized monopole
equation \cite{Klemm:2020mfp}.
If $\Sigma$ were constant (which can always be achieved by a Weyl rescaling \eqref{weylresc}, under which 
$\Sigma\mapsto e^{-\Omega}\Sigma$), \eqref{genmoneq} would boil down to $dh=\star h\Sigma$, which
is the self-duality condition \cite{Townsend:1983xs} in three dimensions. We can thus regard \eqref{genmoneq} as a conformally invariant generalization of the three-dimensional self-duality condition. 
We will further elaborate on this point in the sequel. Notice that, in that case, \eqref{QQtrel} would yield
\begin{equation}\label{relQQt}
\tilde{Q}_\mu = -\frac12 Q_\mu\,.
\end{equation}

Setting $\Sigma$ constant above has the same effects as considering the solutions \eqref{rconst}
(we discard the trivial case $R=0$). Then, $f(R)$ is constant and we can write
\begin{equation}\label{fprimeconstant}
f'(R) = C_0\,,
\end{equation}
where $C_0$ is an arbitrary constant. 
Thus one has $\partial_\lambda f' = 0$, and
\eqref{dff} leads to \eqref{relQQt}, so that we are left with just one independent nonmetricity vector.
Using \eqref{rconst}, \eqref{eq:EW3d} becomes
\begin{equation}\label{eqnewfin}
R_{(\mu\nu)} = \frac{c_k}3 g_{\mu\nu}\,.
\end{equation}
Then, inserting \eqref{fprimeconstant} and \eqref{relQQt} into \eqref{zetaformfin} and \eqref{homotfin},
one obtains
\begin{align}
& \zeta_{\rho\sigma} = \frac{3 C_0}{8\kappa^2}\varepsilon_{\rho\sigma\tau} Q^\tau\,, \label{bpqc2} \\
& \hat{R}_{\rho\sigma} = \frac{C_0\mu}8\varepsilon_{\rho\sigma\tau} Q^\tau\,. \label{rhatc2}
\end{align}
Note that \eqref{rhatc2} can be dualized as
\begin{equation}\label{dual}
-\frac{C_0\mu}4 Q_\alpha = \varepsilon_{\alpha\rho\sigma}\partial^\rho Q^\sigma\,,
\end{equation}
which implies
\begin{equation}\label{covdq}
\tilde{\nabla}^\mu Q_\mu =0\,,
\end{equation}
where $\tilde{\nabla}$ denotes the Levi-Civita covariant derivative.
Using \eqref{relQQt}, the nonmetricity tensor \eqref{nmform} becomes
\begin{equation}\label{schnm}
Q_{\lambda\mu\nu} = \frac12 Q_\lambda g_{\mu\nu} - \frac14 Q_\mu g_{\lambda\nu} - \frac14 Q_\nu 
g_{\lambda\mu}\,,
\end{equation}
and, in particular, we have
\begin{equation}\label{nmsymm}
Q_{(\lambda\mu\nu)} = 0\,.
\end{equation}
Consequently, \eqref{connint} boils down to
\begin{equation}\label{schconn}
{\Gamma^\lambda}_{\mu\nu} = \tilde{\Gamma}^\lambda_{\phantom{\lambda}\mu\nu} - \frac12
g_{\mu\nu} Q^\lambda + \frac12 {\delta_{(\nu}}^\lambda Q_{\mu)} = {\tilde{\Gamma}^\lambda}_{\mu\nu}
- g^{\lambda\rho} Q_{\rho\mu\nu}\,.
\end{equation}
\eqref{schconn} corresponds thus to a Schr\"{o}dinger connection \eqref{schconndef} with \eqref{zqident}
and nonmetricity given by \eqref{schnm}. Moreover, \eqref{dual} implies that in the present case the
nonmetricity vector $Q_\mu$ is selfdual \cite{Townsend:1983xs}.

\subsection{Self-duality in three dimensions and inhomogeneous Maxwell equations}

In what follows, we discuss some interesting consequences of our theory arising from the self-duality
relation \eqref{dual}.

Let us first recall that the authors of \cite{Townsend:1983xs} showed that for spacetime dimension
$n=4k-1$, $k=1,2,3,\ldots$, one may take the `square root' of the Proca equation for a massive
antisymmetric tensor field. The result is a selfdual field; \eqref{dual} corresponds to the case $k=1$
and can be rewritten as
\begin{equation}\label{dualrewr}
Q_\mu = \frac1{2M}\varepsilon_{\mu\nu\rho}\mathcal{F}^{\nu\rho}\,,
\end{equation}
where $M=-C_0\mu/4$ is interpreted as a mass parameter that depends, in particular, on the
Chern-Simons coupling $\mu$, and we defined the field strength
\begin{equation}\label{defF}
\mathcal{F}_{\mu\nu} = 2\tilde{\nabla}_{[\mu} Q_{\nu]} = 2\partial_{[\mu} Q_{\nu]} = 2\hat{R}_{\mu\nu}\,.
\end{equation}
The dual form of \eqref{dualrewr} reads
\begin{equation}\label{dualFdual}
\mathcal{F}_{\alpha\beta} = M\varepsilon_{\mu\beta\alpha} Q^\mu\,.
\end{equation}
One shews that \eqref{dualrewr} implies \eqref{covdq}, together with the covariant Proca equation
\begin{equation}\label{procaQ}
\tilde{\nabla}^\mu\mathcal{F}_{\mu\nu} - M^2 Q_\nu =0\,.
\end{equation}
Notice that, using \eqref{dualFdual}, \eqref{procaQ} can also be written in the form
\begin{equation}\label{procaQnew}
\partial^\mu\mathcal{F}_{\mu\nu} - \frac12 g^{\alpha\beta}\partial^\mu g_{\alpha\beta}
\mathcal{F}_{\mu\nu} - M^2 Q_\nu = 0\,.
\end{equation}
Thus, the vector $Q_\mu$ could be interpreted as a massive photon gauge field. 

Furthermore, \eqref{procaQ} corresponds to an inhomogeneous electromagnetic wave equation which
follows from the inhomogeneous Maxwell equations.
In this context, we find that $Q_\mu$ results to be massless and source of itself. In order to see this 
explicitly, let us consider the following wave equation, implied by \eqref{procaQ} and \eqref{covdq}:
\begin{equation}\label{weQ}
\left(\square - M^2\right) Q^\mu = 0\,,
\end{equation}
where $\square\equiv\tilde{\nabla}_\alpha\tilde{\nabla}^\alpha$.
On the other hand, the inhomogeneous electromagnetic wave equation for $Q^\mu$ in the gauge 
\eqref{covdq} reads
\begin{equation}\label{inmax}
\square Q^\mu = \mu_0 J^\mu\quad\Leftrightarrow\quad\square Q^\mu = \tilde{\nabla}_\rho 
\mathcal{F}^{\rho\mu}\,,
\end{equation}
where $\mu_0$ is the vacuum permeability and $J^\mu$ the current.
Now plug \eqref{defF} and \eqref{dualFdual} into \eqref{weQ} to find
\begin{equation}\label{msdfin}
\square Q^\mu = \frac M2\varepsilon^{\mu\rho\sigma}\mathcal{F}_{\rho\sigma} = \tilde{\nabla}_\rho
\left( M\varepsilon^{\mu\rho\sigma} Q_\sigma\right) = \tilde{\nabla}_\rho\mathcal{F}^{\rho\mu}\,.
\end{equation}
We see that \eqref{msdfin} coincides with \eqref{inmax}, and we can also write
\begin{equation}
J^\mu\equiv\frac1{\mu_0}\tilde{\nabla}_\rho\mathcal{F}^{\rho\mu} = \frac M{2\mu_0}
\varepsilon^{\mu\rho\sigma}\mathcal{F}_{\rho\sigma} = \frac{M^2}{\mu_0} Q^\mu\,.
\end{equation}
The source current $J^\mu$ is covariantly conserved, $\tilde{\nabla}_\mu J^\mu=0$.
Nevertheless, by this last equation we cannot directly define a globally conserved charge, since we need
a local conservation law to do this. On the other hand, considering \eqref{procaQnew}, we deduce that the 
total current (source current plus self-current)
\begin{equation}\label{totcurr}
j^\mu = J^\mu + \frac1{2\mu_0} g^{\alpha\beta}\partial_\rho g_{\alpha\beta}\mathcal{F}^{\mu\rho}
\end{equation}
is locally conserved, $\partial_\mu j^\mu=0$.

Concluding, \eqref{procaQ}, which involves a massive photon, can also be interpreted as the
inhomogeneous Maxwell equations, with corresponding wave equation, where the photon is source of
itself, due to the self-duality relation \eqref{dual}. Gauge invariance of the connection and of the inhomogeneous Maxwell equations follows from self-duality.

\section{Discussion}\label{discussion}

We presented a three-dimensional metric affine theory of gravity whose field equations lead, considering
the particular solution with constant scalar curvature, to a connection introduced by Schr\"{o}dinger in
the 1940s. Although involving nonmetricity, the latter preserves the length of vectors under parallel
transport.
Furthermore, we obtained a self-duality relation for the nonmetricity vector $Q_\mu$ leading to a Proca 
equation which may also be interpreted as inhomogeneous Maxwell equation. Gauge invariance follows from 
self-duality and we can conclude that, in our framework, the inhomogeneous Maxwell equations emerge
from affine geometry, i.e., from a purely gravitational setup.

Let us mention that similar results were obtained in \cite{Klemm:2020mfp} (where the
authors presented for the first time an action principle for the Einstein-Weyl equations in three dimensions),
where the Weyl vector is self-dual.
We also remark that the Chern-Simons and Lagrange multiplier terms in \eqref{actiontot} break the
projective invariance of the connection, which allows for consistent matter couplings (c.f. the discussion
in \cite{Klemm:2020mfp}).

The model and the results presented in this paper not only appear to be relevant under geometrical and 
physical perspectives but they also aim to highlight some peculiar and intriguing features of the 
Schr\"{o}dinger connection that have been somewhat overlooked in the past literature.
Future developments could consist in studying possible applications of our results in the classification of 
supersymmetric near-horizon geometries along the lines of \cite{Dunajski:2016rtx, Klemm:2019izb}, and
in the phenomenology related to dark matter and dark energy. It remains to be seen if our model can be 
extended to higher dimensions.

\bigskip

\section*{Acknowledgements}


This work was supported partly by INFN and by MIUR-PRIN contract 2017CC72MK003. L.~R.~would like to 
thank A.~Gamba and F.~Dolcini for financial support.

\end{document}